\documentclass[prl,nobibnotes,twocolumn,aps,longbibliography,superscriptaddress,notoc]{revtex4-1}
\usepackage{bm}
\usepackage{graphicx}
\usepackage{amssymb}
\usepackage{amsmath}
\usepackage{eufrak}
\usepackage{color}
\usepackage[utf8]{inputenc}
\usepackage{ulem}
\usepackage[unicode=true,colorlinks=true,citecolor=blue,urlcolor=blue]{hyperref}
\usepackage{dsfont}

\newcommand{\Tr}{\mathop{\rm Tr}}

\newcommand{\e}{\mathrm{e}}

\let\ifr\i
\renewcommand{\i}{{\rm i}}
\renewcommand{\d}{\mathrm d}
\renewcommand{\emph}{\textit}
\newcommand{\braket}[1]{\left\langle #1 \right\rangle}

\usepackage{chngcntr}
\newcommand{\enquote}{}
\newcommand{\nix}[1]{}
\let\oldsec\section
\renewcommand{\section}[1]{\textit{#1}---}

\begin{document}

\title{Cooling and heating nuclear spins by strongly localized electrons}

\author{D.~S.~Smirnov}
\email[Electronic address: ]{smirnov@mail.ioffe.ru}
\affiliation{Ioffe Institute, 194021 St. Petersburg, Russia}
\author{K.~V.~Kavokin}
\affiliation{Spin Optics Laboratory, Saint Petersburg State University, 198504 Saint Petersburg, Russia}

\begin{abstract}
  The concept of nuclear spin temperature has been a cornerstone of the theory of dynamic nuclear spin polarization by electrons in various semiconductor structures for decades. Still, it is not always applicable to strongly localized electrons due to their long spin correlation times. This motivated the use of the oversimplified central spin model for the description of the nuclear spin dynamics in quantum dots. Here, we present a microscopic theory that bridges the gap between these two approaches by describing the nuclear spin thermodynamics for systems with long electron spin correlation times. Importantly, our theory predicts that efficient nuclear spin cooling by strongly localized electrons requires an external magnetic field by far exceeding the local field of nuclear spin-spin interaction, and that the time of the nuclear spin heating by unpolarized electrons may change by several orders of magnitude depending on the magnetic field.
\end{abstract}

\maketitle{}

Dynamic spin polarization of lattice nuclei by optically oriented electrons in semiconductors is known for over 50 years since the experiments by Lampel~\cite{lampel68} in silicon and by Ekimov and Safarov in III-V direct-gap semiconductors~\cite{PismaZhETF.15.257}. Technically, this phenomenon is a specific case of the Overhauser effect~\cite{PhysRev.92.411}, with the deviation of the mean spin of electrons from thermal equilibrium by optical pumping with circularly polarized light. The return of the electron spin ensemble to the equilibrium is accompanied by the transfer of angular momentum to the nuclear spin system via the hyperfine interaction, while the excess energy goes to the lattice via the electron-phonon interaction. As optical orientation can provide high spin polarization of electrons, nuclear spins can be efficiently polarized even in weak external magnetic fields. The theory by Dyakonov and Perel~\cite{dp74} was proposed immediately after those experimental discoveries and served as a basis for the studies of electron-nuclear spin dynamics in semiconductors ever since~\cite{book_Glazov}.

The main approximation made within the Dyakonov-Perel theory is that of a short time during which a localized electron interacts with the nuclei at the given localization site. It assumes that this so-called correlation time is much shorter than the period of the electron spin precession in the effective magnetic field created by randomly oriented spins of these nuclei via their hyperfine interaction with the electron. This approximation holds for a vast variety of semiconductor structures, including bulk crystals, epitaxial layers and quantum wells~\cite{dyakonov_book,book_Glazov}.

The list of structures where it does not hold includes, first of all, various types of quantum dots~\cite{A.Greilich09282007,Reilly817,xu09,maletinsky2009breakdown,Gangloff2021,cai2024nuclearspin}. Recently it was complimented by lead halide perovskites, where strongly localized charge carriers (both electrons and holes) demonstrate spin dynamics incompatible with short correlation time~\cite{kirstein2023squeezed,kudlacik2024optical}. Several theoretical works~\cite{merkulov02,PhysRevLett.88.186802,PhysRevB.81.115107,Glazov_hopping,PhysRevLett.126.216804} devoted to the spin dynamics of charge carriers having long spin correlation time, were based on the central spin model. However, since the possibility to account for the nuclear spin-spin interactions within this approach is very limited, a theory of the dynamic nuclear spin polarization valid throughout the range of electron correlation times (and applied magnetic fields) is still lacking. We aim this work at filling this gap.

To analyse dynamic polarization of the nuclear spin system (NSS), we apply the spin temperature concept (in line with Ref.~\onlinecite{dp74}). It is known to correctly describe the nuclear spin dynamics in a variety of solids~\cite{goldman1970spin}, and was recently experimentally verified in semiconductor structures both for high~\cite{Chekhovich_constants} and low~\cite{PhysRevB.97.041301} magnetic fields. The spin temperature concept is based on the experimental fact that spin-lattice relaxation times of nuclei in solids (at least at cryogenic temperatures of the crystal lattice) are orders of magnitude longer than characteristic times of spin-spin interactions of neighboring nuclei. Nuclear spin-spin interactions are usually dominated by the dipole-dipole magnetic interaction, and therefore are anisotropic and (considering many nuclear spins involved) in a sense random. They rapidly destroy all the spin correlations but leave the total energy of the NSS unchanged. For this reason, beyond the spin-spin time $T_{2,N}$ the NSS density matrix has only diagonal elements proportional to Boltzmann exponentials with a so-called nuclear spin temperature $\theta_N$. It defines all the quantities pertinent to the NSS and, in particular, the mean nuclear spin in the magnetic field $B$ applied along the $z$ axis:
\begin{equation}
  \label{eq:P_N}
  \braket{I_z}=\frac{I(I+1)}{3k_B\theta_B}\hbar\gamma_NB,
\end{equation}
where $\gamma_N$ is the nuclear gyromagnetic ratio, $k_B$ is the Boltzmann constant, and $I$ is the spin of the nuclear species comprising the NSS. Hereafter we use the high-temperature approximation.

\begin{figure}[htb]
\centering
\includegraphics[width = 0.9\linewidth]{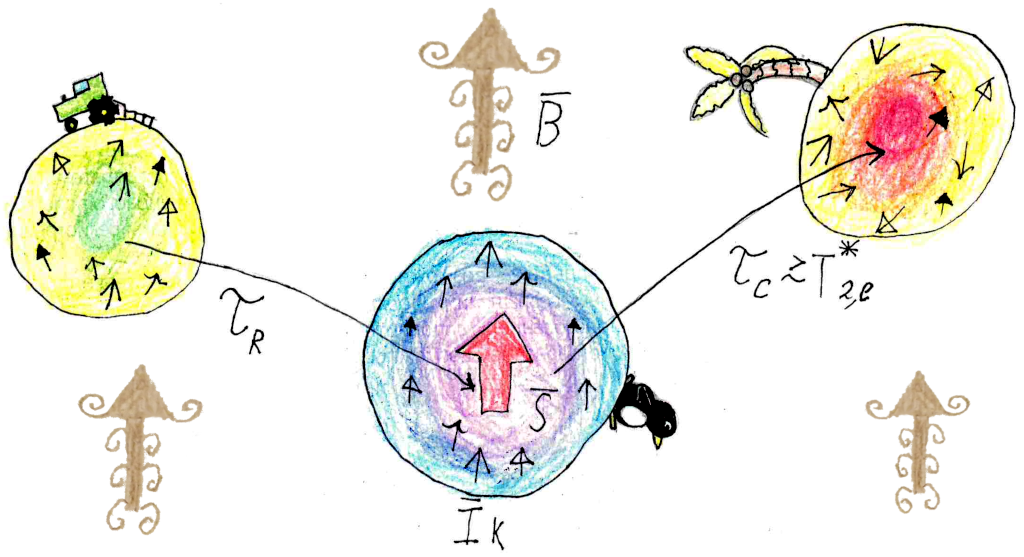}
\caption{\label{model}
  Impression of cooling and heating of the nuclear spins $\bm I_k$ (small black arrows) in the quantum dots by strongly localized electron spin $\bm S$ (large red arrow) in external magnetic field $\bm B$ (curly brown arrows). Colder nuclear spins are better aligned along $\bm B$. Electron hopping time to the given quantum dot $\tau_R$ and away from it $\tau_c$ can be comparable to or even larger than the electron spin dephasing time $T_{2,e}^*$.
}
\end{figure}

Nuclear spin temperature establishes in the dynamic equilibrium under optical pumping as a result of the energy exchange with the electrons~\cite{OptOrCh5}. We assume that the electrons with the mean spin $\braket{\bf{S}}$ come to the given localization site (quantum dot) with the periodicity $\tau_R$ and stay during a correlation time $\tau_c$ so that the filling factor of the quantum dots equals $f=\tau_c/\tau_R$, see Fig.~\ref{model}. The electron spin in the quantum dot rotates with the typical period $T_{2,e}^*\sim\hbar/(\sqrt{N}a)$ in the Overhauser field of mostly randomly oriented $N$ nuclear spins, where $a$ is the hyperfine coupling constant with one nucleus. The previous theories where limited to the case of short $\tau_c$ compared to $T_{2,e}^*$, while we allow the relation between them to be arbitrary.

The balance of the energy fluxes from and to the NSS reads
\begin{equation}
  \label{eq:balance}
  \frac{f\braket{S_z}}{T_{1,e}^N}\hbar\gamma_N B=q_Z+q_{SS},
\end{equation}
Here $\braket{S_z}=\left|\braket{\bm S}\right|\cos\theta$ is the projection of the electron mean spin on the external field with $\theta$ being the angle between the optical and $z$ axes, and $T_{1,e}^N$ is the electron spin relaxation time due to the hyperfine interaction. We neglect the other sources of the electron spin relaxation, which are less important at low temperatures and magnetic fields~\cite{PhysRevB.64.125316,PhysRevB.66.161318}. The left-hand side of Eq.~\eqref{eq:balance} represents the rate of changing the Zeeman energy of nuclei due to the spin influx from electrons. The right-hand side is a sum of energy fluxes due to the warm-up of the Zeeman ($q_Z$) and spin-spin ($q_{SS}$) nuclear reservoirs by the electron spin fluctuations. If the spin polarization of electrons $\braket{\bm S}$ is small, then $q_Z$ and $q_{SS}$ are independent of it.

The spin relaxation time of the hopping electrons, induced by nuclei, for arbitrary correlation time reads~\cite{Glazov_hopping,PRC}
\begin{equation}
  \label{eq:T_eN}
  T_{1,e}^N=\tau_c\braket{\frac{\Omega_{e,x}^2+\Omega_{e,y}^2}{\Omega_e^2+1/\tau_c^2}}^{-1},
\end{equation}
where $\bm\Omega_e$ is the electron spin precession frequency in the sum of the Overhauser and external magnetic fields and the angular brackets denote averaging over the nuclear spin fluctuations.

The flux $q_Z$ can be found from the principle of detailed balance: in the absence of the dipole-dipole interaction ($q_{SS}=0$), flip-flop transitions result in equilibration of the population ratio of electron spin levels with the population ratios of any nuclear spin levels with the angular momentum $I_z$ differing by one. This leads to the relation between mean spins of electrons and nuclei $\braket{I_z}=\braket{S_z}\cdot 4I(I+1)/3$, which implies that
\begin{equation}
  \label{eq:q_Z}
  q_Z=\frac{3f\braket{I_z}}{4I(I+1)T_{1,e}^N}\hbar\gamma_N B.
\end{equation}

The nuclear spin-spin reservoir is heated by pulses of the Knight field. It rotates the nuclear spins by some angle. Though this angle is virtually the same for adjacent spins, the dipole-dipole energy changes, since it depends on the orientation of the two interacting nuclear spins with respect to the straight line connecting the nuclei.

The general expression for $q_{SS}$ can be obtained in the important limit of $\tau_c\ll\hbar/a$, when the nuclear spin rotation angles are small (but $\tau_c$ can be still greater than $T_{2,e}^*$). In this limit, for any pair of the collective nuclear spin states $n$ and $m$, the transition probability $w_{nm}$ is proportional to the squared product of the interaction strength and the pulse duration, i.e. $w_{nm}\propto a^2 \tau_c^2/\hbar^2$. The difference of the corresponding populations $p_n-p_m=\beta(E_m-E_n)$ is determined by their energies $E_{n,m}$ and the inverse spin temperature $\beta=1/(k_B\theta_N)$. Therefore, the energy changing rate is
\begin{multline}
  \label{eq:q_SS}
  q_{SS}=\beta\tau_R^{-1}\frac{1}{2}\sum_{n,m}w_{nm}(E_n-E_m)^2 \\
  =\Phi N \frac{I(I+1)}{3}(\gamma_N B_L)^2a^2\tau_c^2\beta/\tau_R,
\end{multline}
where $B_L$ is the local field determined by the nuclear spin-spin interactions~\cite{goldman1970spin}, and the exact expression for the dimensionless coefficient $\Phi$ is obtained below by the spin density matrix method. From qualitative considerations, one can conclude that $\Phi$ depends on the electron spin correlation time, since if $\tau_c$ is short, all three electron spin components are equally efficient, while at long $\tau_c$, the components transverse to the nuclear fluctuation are rapidly averaged out by the precession around the fluctuation field, and only the longitudinal component remains efficient during all the interaction time. For this reason, in the transition from short to long correlation time, $\Phi$ diminishes 3 times and then does not change with further increase of $\tau_c$.

\begin{figure*}
  \centering
  \includegraphics[width=\linewidth]{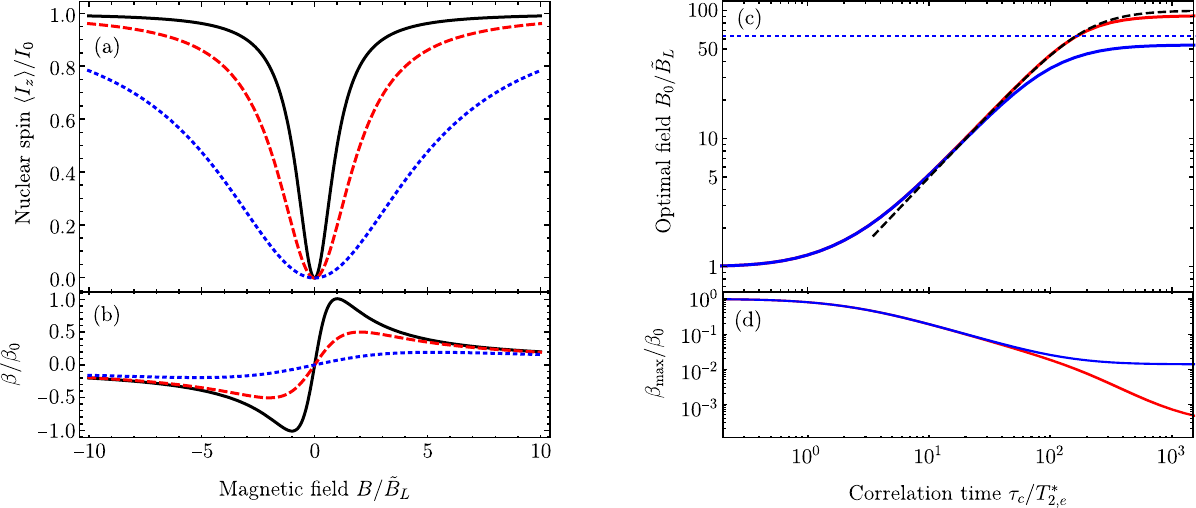}
  \caption{Dependences on the magnetic field of (a) the nuclear spin polarization and (b) the inverse nuclear spin temperature for $\tau_c\to 0$ (black solid line), $\tau_c=3T_{2,e}^*$ (red dashed line), and $\tau_c=10T_{2,e}^*$ (blue dotted line). Dependences on the electron correlation time $\tau_c$ of (c) the optimal field $B_0$ at which the maximum of $\beta$ equal to $\beta_0$ is reached and (d) of $\beta_0$. The parameters are $\Delta_B=100\tilde{B}_L$ and $N_I=10^6$ (red lines) and $N_I=10^4$ (blue lines). Black dashed line in (c) is calculated after Eq.~\eqref{eq:B0_anal}.}
  \label{fig:combo}
\end{figure*}

Substituting Eqs.~\eqref{eq:q_Z} and~\eqref{eq:q_SS} in Eq.~\eqref{eq:balance} and making use of Eq.~\eqref{eq:P_N} we obtain the inverse spin temperature established under dynamic polarization by electrons:
\begin{equation}
  \label{eq:beta}
  \beta=\frac{4\braket{S_z}B/(\hbar\gamma_N)}{B^2+\Phi\frac{T_{1,e}^N}{\tau_c}(\tau_c/T_{2,e}^*)^2B_L^2},
\end{equation}
where we have set $T_{2,e}^*=\hbar/(a\sqrt{N_I})$ with $N_I=4NI(I+1)/3$.

In particular, in the limit of short correlation time, Eq.~\eqref{eq:T_eN} reduces to $T_{1,e}^N=1/\left(\braket{\Omega_{e,x}^2+\Omega_{e,y}^2}\tau_c\right)$, so Eq.~\eqref{eq:beta} reproduces the classical result of the Dyakonov-Perel theory, to which end one should put $\Phi=3/2$~\cite{dp74,abragam,OptOrCh2}.

In the opposite limit of long correlation time, $\Phi$ then becomes equal to $1/2$. The electron spin relaxation time by nuclei in this limit equals $3\tau_c/2$, which reflects the fact that the electron loses 2/3 of its spin polarization with each coming to a new localization center. Thus, in the limit of long correlation time, the inverse nuclear spin temperature is
\begin{equation}
  \label{eq:betalong}
  \beta=\frac{4\braket{S_z}B/(\hbar\gamma_N)}{B^2+\left(\tau_c/T_{2,e}^*\right)^2B_L^2/4}.
\end{equation}
This expression reveals our main result: For $\tau_c\gg T_{2,e}^*$, magnetic fields much larger than $B_L$ are needed to efficiently cool nuclear spins. This happens because electrons cool the nuclei during the time $\sim T_{2,e}^*$ only, but heat the nuclear spin-spin reservoir during much longer time $\tau_c$.

In Ref.~\onlinecite{supp}, we support these general considerations with microscopic calculations of the nuclear spin dynamics for the following Hamiltonian:
\begin{equation}
  \label{eq:interaction}
  \mathcal H=a\bm J\bm S+\hbar\gamma_e\bm{SB}-\hbar\gamma_N\bm{JB}+\mathcal H_{SS},
\end{equation}
where $\bm J=\sum_{n=1}^N\bm I_n$ is the total spin of all $N$ nuclei $\bm I_n$ in the quantum dot, $\gamma_e$ is the electron gyromagnetic ratio, and $\mathcal H_{SS}$ describes the nuclear spin-spin (dipole-dipole) interactions. The hyperfine interaction is assumed to be homogeneous (the ``box'' approximation). The condition for the establishment of the nuclear spin temperature in this model is provided also in Ref.~\onlinecite{supp}.

Making use of the separation of the time scales of electron and nuclear spin dynamics we find
\begin{equation}
  \label{eq:Phi_ans}
  \Phi=\left<\frac{1}{1+\Omega_e^2\tau_c^2}\right>+\frac{1}{2}.
\end{equation}
One can see that it equals $3/2$ for short correlation times, and decreases $3$ times with increase of $\tau_c$ in agreement with the above qualitative considerations.

Eqs.~\eqref{eq:T_eN} and~\eqref{eq:Phi_ans} allow one to calculate the nuclear spin temperature from Eq.~\eqref{eq:beta}, for arbitrary correlation time assuming the nuclear spin polarization much smaller than $1/\sqrt{N}$. The dependences of the mean spin of nuclei (normalized to $I_0=\frac{4}{3}I(I+1)\braket{S_z}$) and inverse spin temperature (normalized to $\beta_0=2\braket{S_z}/(\hbar\gamma_N\sqrt{3}B_L)$) on the external magnetic field are shown in Fig.~\ref{fig:combo}(a,b). The black lines correspond to the classical limit of short correlation time. One can see, that the stronger the field, the larger the nuclear spin polarization, while the inverse nuclear spin temperature has two extrema as a function of $B$. With growing $\tau_c$, these dependences broaden. The maximum of $\beta$ decreases, but the nuclear spin still saturates at the same value.

Generally, $\beta$ is well described by the expression $2BB_0\beta_{\rm{max}}/(B^2+B_0^2)$ with the maximum of $\beta_{\rm max}$ reached at $B_0$. We plot these parameters in Fig.~\ref{fig:combo}(d,c) for $N_I=10^6$ (red lines) and $10^4$ (blue lines) nuclear spins. At short correlation times, $B_0$ equals the local field $\tilde{B}_L=\sqrt{3}B_L$, as expected~\cite{dp74}. For a large number of nuclei, $N_I\gg(T_{2,e}^*\gamma_eB_L)^{-2}$, and at long correlation times, $\tau_c\gg T_{2,e}^*$, we obtain~\cite{supp}
\begin{equation}
  \label{eq:B0_anal}
  B_0=\frac{1/T_{2,e}^*}{\sqrt{\gamma_e^2+4/(\tau_c\tilde{B}_L)^2}}.
\end{equation}
This dependence is shown by the black dashed line in Fig.~\ref{fig:combo}(c) and agrees well with the red curve. In particular, $B_0$ grows linearly with $\tau_c$ and saturates at $1/(\gamma_eT_{2,e}^*)$. For smaller number of nuclei, $B_0$ saturates at $\sqrt{2N_I/5}\tilde{B}_L$ [blue dotted line in Fig.~\ref{fig:combo}(c)]. This saturation corresponds to the maximum possible value of $q_{SS}$, $q_{SS,max}\sim N(\hbar\gamma_N B_L)^2\beta/\tau_R$, achieved when the nuclear spins are completely reoriented as a result of interaction with the electron. In both cases, with increase of $\tau_c$, the optimal field increase by a large factor from tens to a hundred, as one can see in Fig.~\ref{fig:combo}(c).

Thus, the main implication of the presented theory for the experiments on optical cooling of the NSS in weak magnetic fields is the unexpected reduction of the dynamic nuclear polarization efficiency in structures with apparently stronger electron-nuclear coupling provided by better localization of electrons. Cooling of nuclear spins with strongly localized electrons turns out to be hindered by intense heating with the same electrons. Application of external magnetic field exceeding many times the local field of the nuclear spin-spin interactions might be needed to overcome the electron-induced nuclear spin warm-up.

It is also worth considering the electron-induced relaxation of the nuclear spin temperature to the lattice temperature ``in the dark'', i.e. after switching off the optical pumping. In this regime, the dynamics of the inverse spin temperature is determined by warming up of Zeeman and spin-spin energy reservoirs:
\begin{equation}
  \dot{\beta}=\frac{q_Z+q_{SS}}{\partial E_{NSS}/\partial\beta},
\end{equation}
where
\begin{equation}
  \label{eq:C}
  -\frac{\partial E_{NSS}}{\partial\beta}=N_I(\hbar\gamma_N/2)^2(B^2+B_L^2)
\end{equation}
is the heat capacity of the NSS~\cite{goldman1970spin}. This leads the following general expression for the electron-induced nuclear spin-lattice relaxation rate:
\begin{equation}
  \label{eq:TE}
  \frac{1}{T_{\mathcal E}}=-\frac{\dot{\beta}}{\beta}=\frac{f}{N_I\left(B^2+B_L^2\right)}\left(\frac{B^2}{T_{1,e}^N}+\frac{\Phi\tau_cB_L^2}{T_{2,e}^{*2}}\right).
\end{equation}
We use the notation $T_{\mathcal E}$ to avoid a confusion with the longitudinal nuclear spin relaxation time $T_{1,N}^e$ which, in principle, may be different.

\begin{figure}
  \includegraphics[width=\linewidth]{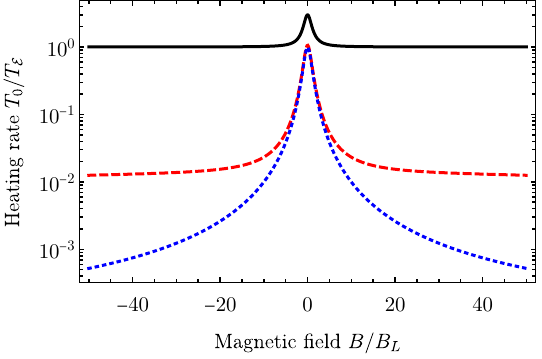}
  \caption{\label{fig:heating} Heating rate calculated after Eq.~\eqref{eq:TE} for $\tau_c\to 0$ (black solid line), $\tau_c=10T_{2,e}^*$ (red dashed line), and $100T_{2,e}^*$ (blue dotted line).
  }
\end{figure}

The heating rate (normalized to $T_0^{-1}=f\tau_c/(2N_IT_{2,e}^{*2})$) is shown in Fig.~\ref{fig:heating} as a function of magnetic field. As well as the dynamic polarization efficiency, the heating rate at short correlation time (black line) follows the classical theory: It has a Lorentzian form and decreases thrice with increase of magnetic field beyond $B_L$. For longer correlation times, the heating rate at zero magnetic field decreases thrice, but most importantly it changes much more than three times with magnetic field.

Such a strong suppression of nuclear spin relaxation ``in the dark'' by magnetic field was indeed observed in Refs.~\onlinecite{Vladimirova2017,Gribakin2024}. But it was explained there by different mechanisms taking into account the nuclear spin diffusion towards impurity centers~\cite{Khutsishvili}.

In conclusion, strongly localized electrons exchange energy with the nuclear spin-spin reservoir the more efficiently, the longer they interact with the nuclear spins. To the opposite, the energy exchange with the Zeeman reservoir is limited by the angular momentum conservation. As a consequence, in external fields of the order of the local fields the dynamic polarization of the nuclear spin system is suppressed, while the spin-lattice relaxation is enhanced. In order to cool the nuclear spins by optical pumping, one should apply magnetic fields exceeding the local field by a large factor determined by the squared product of spin correlation time of electrons and their hyperfine energy due to nuclear spin fluctuations.

\let\oldaddcontentsline\addcontentsline
\renewcommand{\addcontentsline}[3]{}
We thank A. D. Smirnov for design and drawing of Fig.~\ref{model}. We acknowledge the Foundation for the Advancement of Theoretical Physics and Mathematics “BASIS.” K.V.K. acknowledges the financial support by the Saint Petersburg State University through Research Grant No. 122040800257-5. Numerical calculation of the nuclear spin temperature by D.S.S. was supported by the Russian Science Foundation Grant No. 23-12-00142.

\renewcommand{\i}{\ifr}


%

\let\addcontentsline\oldaddcontentsline
\makeatletter
\renewcommand\tableofcontents{%
    \@starttoc{toc}%
}
\makeatother
\renewcommand{\i}{{\rm i}}

\onecolumngrid
\vspace{\columnsep}
\begin{center}
\makeatletter
{\large\bf{Supplemental Material to\\``\@title''}}
\makeatother
\end{center}
\vspace{\columnsep}

The Supplementary Material includes the following topics:

\hypersetup{linktoc=page}
\tableofcontents
\vspace{\columnsep}
\twocolumngrid

\counterwithin{figure}{section}
\renewcommand{\section}[1]{\oldsec{#1}}
\renewcommand{\thepage}{S\arabic{page}}
\renewcommand{\theequation}{S\arabic{equation}}
\renewcommand{\thefigure}{S\arabic{figure}}
\renewcommand{\bibnumfmt}[1]{[S#1]}
\renewcommand{\citenumfont}[1]{S#1}

\setcounter{page}{1}
\setcounter{section}{0}
\setcounter{equation}{0}
\setcounter{figure}{0}

\section{S1. Derivation of heat fluxes}
\label{sec:S1}

\subsection{General formalism}

Nuclear spin Hamiltonian consists of two terms
\begin{equation}
  \mathcal H_n=\mathcal H_Z+\mathcal H_{SS},
\end{equation}
where $\mathcal H_Z=-\hbar\gamma_NBJ_z$ is the Zeeman Hamiltonian with $\gamma_N$ being the nuclear gyromagnetic ratio, $B$ being external magnetic field applied along the $z$ axis, and $\bm J=\sum_{i=1}^N\bm I_i$ being the total nuclear spin in the quantum dot composed of $N$ individual nuclear spins $\bm I_i$ (their absolute values $I$ are all equal), and $\mathcal H_{SS}$ is the Hamiltonian of the nuclear spin-spin interactions. Before arrival of an electron to the quantum dot, its density matrix reads
\begin{equation}
  \label{eq:rho0_S}
  \rho_0=1-\beta\mathcal H_n,
\end{equation}
where $\beta$ is an inverse nuclear spin temperature measured in the inverse energy units and the normalization $\Tr\rho_0=1$ is implicitly assumed.

When an electron with the spin $\bm S$ comes to the quantum dot, additional terms appear in the total Hamiltonian [Eq.~\eqref{eq:interaction} in the main text]:
\begin{equation}
  \label{eq:H_S}
  \mathcal H=\mathcal H_n+\hbar\gamma_eBS_z+\mathcal H_{hf}
\end{equation}
with $\gamma_e$ being the electron gyromagnetic ratio and
\begin{equation}
  \mathcal H_{hf}=a\bm J\bm S
\end{equation}
being the Hamiltonian of the homogeneous hyperfine interaction (``box approximation'') with a hyperfine coupling constant $a$.

Then the evolution of the nuclear spin density matrix is described by
\begin{equation}
  \label{eq:rho_t}
  \rho(t)=\e^{-\i\mathcal H t/\hbar}\rho_0\e^{\i\mathcal H t/\hbar}.
\end{equation}
In particular, energy of the nuclear spin system is
\begin{equation}
  \label{eq:En}
  E_n(t)=\Tr\left[\mathcal H_n\rho(t)\right].
\end{equation}
An electron stays in the quantum dot for a typical time $\tau_c$, so the average change of the energy is given by
\begin{equation}
  \label{eq:dE_av}
  \braket{\Delta E_n}=\int\limits_0^\infty E_n(t)\e^{-t/\tau_c}\frac{\d t}{\tau_c}-E_n(0).
\end{equation}

The energy flux can be derived from Eq.~\eqref{eq:dE_av} as $q=\braket{\Delta E_n}/\tau_R$, where $\tau_R$ is the typical electron arrival time to the quantum dot. The different terms in $q$ are introduced in Eq.~\eqref{eq:balance} in the main text. Let us calculate them explicitly.

\subsection{Zeeman reservoir}

Heating and cooling of the nuclear Zeeman reservoir takes place at the time scale of the electron spin dephasing time $T_{2,e}^*$. It is described by Eq.~\eqref{eq:dE_av} neglecting $\mathcal H_{SS}$. Then Eq.~\eqref{eq:H_S} reduces to the central spin Hamiltonian, which is already well studied~\cite{S_book_Glazov}.

In particular, the electron spin dynamics is described by
\begin{equation}
  \label{eq:dS}
  \frac{\d\bm S}{\d t}=\bm\Omega_e\times\bm S,
\end{equation}
where $\bm\Omega_e=\bm\Omega_B+\bm\Omega_N$ is the electron spin precession frequency contributed by the external magnetic field $\bm\Omega_B=\gamma_e\bm B$ and nuclear Overhauser field $\bm\Omega_N=a\bm J/\hbar$. At the timescale $t\ll\hbar/a$ the nuclear spins are almost frozen, so the electron spin precesses with almost constant but random frequency~\cite{S_merkulov02}. The corresponding solution of Eq.~\eqref{eq:dS}, $\bm S_0(t)$, reads
\begin{multline}
  \label{eq:S0}
  \bm S_0(t)=\left[\bm S(0)\bm n\right]\bm n+\left\{\bm S(0)-\left[\bm S(0)\bm n\right]\bm n\right\}\cos\left(\Omega_e t\right)\\
  +\left[\bm S(0)\times\bm n\right]\sin\left(\Omega_e t\right),
\end{multline}
where $\bm n=\bm\Omega_e/\Omega_e$.

The longitudinal electron spin relaxation time $T_{1,e}^N$ can be found from Eq.~\eqref{eq:S0} by considering $\bm S(0)$ parallel to the $z$ axis and averaging over nuclear spin distribution~\eqref{eq:rho0_S}:
\begin{equation}
  \braket{S_{0,z}(t)}=S(0)\braket{\frac{\Omega_{e,z}^2}{\Omega_e^2}+\frac{\Omega_{e,\perp}^2}{\Omega_e^2}\cos\left(\Omega_et\right)},
\end{equation}
where $\Omega_{e,\perp}^2=\Omega_{e,x}^2+\Omega_{e,y}^2$. Now the time $T_{1,e}^N$ can be found from
\begin{equation}
  \label{eq:T1e_def}
  \frac{S(0)}{T_{1,e}^N}=\frac{1}{\tau_c}\int\left[S(0)-\braket{S_{0,z}(t)}\right]\e^{-t/\tau_c}\frac{\d t}{\tau_c}
\end{equation}
and reads~\cite{S_Glazov_hopping,S_PRC}
\begin{equation}
  \label{eq:T1e}
  \frac{1}{T_{1,e}^N}=\braket{\frac{\Omega_{e,\perp}^2\tau_c}{1+\Omega_e^2\tau_c^2}}.
\end{equation}
For short electron correlation time, this gives
\begin{equation}
  T_{1,e}^N=2T_{2,e}^{*2}(1+\Omega_B^2\tau_c^2)/\tau_c,
\end{equation}
where
\begin{equation}
  T_{2,e}^{*}=\frac{\hbar}{a\sqrt{N_I}}
\end{equation}
is the typical period of electron spin precession in the Overhauser field fluctuations with $N_I=4NI(I+1)/3$, which reduces to $N$ for $I=1/2$. In the opposite limit of long correlation time, $\tau_c\gg T_{2,e}^*$, the longitudinal electron spin relaxation time is
\begin{equation}
  T_{1,e}^N=\tau_c/\braket{\Omega_{e,\perp}^2/\Omega_e^2},
\end{equation}
so it is of the order of $\tau_c$ in small magnetic fields and increases as $B^2$ in strong magnetic fields. This dependence can be well approximated (with accuracy of a few percent~\cite{S_petrov08,S_book_Glazov}) by a Lorentzian:
\begin{equation}
  \frac{1}{T_{1,e}^N}=\frac{2}{3\tau_c}\frac{1}{1+\Omega_B^2T_{2,e}^{*2}}.
\end{equation}
These expressions describe the left hand side of Eq.~\eqref{eq:balance} as well as Eq.~\eqref{eq:q_Z} in the main text.

This can be shown more rigorously by considering equations of the nuclear spin dynamics, which follows from Eq.~\eqref{eq:H_S}:
\begin{equation}
  \label{eq:dI}
  \frac{\d\bm J}{\d t}=\bm\omega_N\times\bm J,
\end{equation}
where $\bm\omega_N=-\gamma_N\bm B+a\bm S/\hbar$ is the nuclear spin precession frequency in the sum of external magnetic and Knight fields. The nuclear Zeeman energy is $E_Z=-\hbar\gamma_NBJ_z$, and from comparison of Eqs.~\eqref{eq:dS} and~\eqref{eq:dI} one can see that $\d J_z/\d t=-\d S_z/\d t$ due to the conservation of the total angular momentum component along $\bm B$ in the central spin model. So the change of $E_Z$ (in the main order in $1/\sqrt{N}$) is
\begin{equation}
  \Delta E_Z(t)=\hbar\gamma_NB\left[S_{0,z}(t)-S(0)\right].
\end{equation}
Now comparing this with Eqs.~\eqref{eq:T1e_def} and~\eqref{eq:dE_av} we obtain the energy flux from the nuclear Zeeman reservoir
\begin{equation}
  \frac{f\braket{S_z}}{T_{1,e}^N}\hbar\gamma_N B
\end{equation}
in agreement with Eq.~\eqref{eq:balance} in the main text. Here $f=\tau_c/\tau_R$ is the average occupancy of the quantum dot with electrons and $\braket{S_z}$ is the average spin of hopping electrons along the $z$ axis. This determines the nuclear spin cooling rate.

Heating of the nuclear Zeeman reservoir takes place in the next order in $1/\sqrt{N}$. To derive it, we consider a correction $\delta\bm S$ to Eq.~\eqref{eq:S0} due to the time dependence of $\bm\Omega_N(t)$. From Eq.~\eqref{eq:dS} one can see that it satisfies
\begin{equation}
  \label{eq:d_delta_S}
  \frac{\d\delta\bm S}{\d t}=\bm\Omega_e(0)\times\delta\bm S+\delta\bm\Omega_N\times\bm S_0,
\end{equation}
where $\delta\bm\Omega_N=\bm\Omega_N(t)-\bm\Omega_N(0)$. It can be found from the equation for the nuclear spin dynamics~\eqref{eq:dI}. Change of the nuclear Zeeman energy takes place at the time scale of $T_{2,e}^*$ when the nuclear spin rotation angle is small. Then $\delta\bm J=\bm J(t)-\bm J(0)$ can be calculated by integrating Eq.~\eqref{eq:dI} neglecting the time dependence of $\bm J$ in the right-hand side. Then we substitute $\delta\bm\Omega_N=(a/\hbar)\delta\bm J$ in Eq.~\eqref{eq:d_delta_S} and take into account smallness of the average electron spin polarization. Further we solve this equation for $\delta\bm S(t)$ making use of Eq.~\eqref{eq:S0}. Finally, the average correction to the change of the electron spin is
\begin{equation}
  \braket{\delta\bm S}=\braket{\int\limits_0^\infty\delta\bm S(t)\frac{\e^{-t/\tau_c}}{\tau_c}\d t}.
\end{equation}
While $\braket{\delta S_{x,y}}=0$, we obtain
\begin{equation}
  \label{eq:delta_S1}
  \braket{\delta S_z}=\braket{\frac{a^2\tau_c^2\left[J_z\left(1+\Omega_{e,z}^2\tau_c^2\right)+\hbar\Omega_{e,z}\Omega_{e,\perp}^2\tau_c^2/a\right]}{2\hbar^2\left(1+\Omega_e^2\tau_c^2\right)^2}}.
\end{equation}
The intermediate expressions are very cumbersome.

The averaging in Eq.~\eqref{eq:delta_S1} should be performed with the nuclear spin density matrix~\eqref{eq:rho0_S}. It corresponds to the distribution function of the Overhauser field
\begin{equation}
  \mathcal F(\bm\Omega_N)=\left(\sqrt{\frac{2}{\pi}}T_{2,e}^*\right)^3\exp\left[-2T_{2,e}^{*2}\left(\bm\Omega_N-\braket{\bm\Omega_N}\right)^2\right]
\end{equation}
with the average Overhauser field $\braket{\bm\Omega_N}=a\braket{\bm J}/\hbar$ along the $z$ axis, where $\braket{\bm J}=N\braket{\bm I}$ is the average nuclear spin determined by the inverse nuclear spin temperature [Eq.~\eqref{eq:P_N} in the main text]
\begin{equation}
  \braket{I_z}=\frac{I(I+1)}{3}\hbar\gamma_NB\beta.
\end{equation}
Then one can show that Eq.~\eqref{eq:delta_S1} is equivalent to
\begin{equation}
  \label{eq:delta_S_ans}
  \braket{\delta S_z}=\frac{\braket{J_z}}{N_I}\braket{\frac{\Omega_{e,\perp}^2\tau_c^2}{1+\Omega_e^2\tau_c^2}}.
\end{equation}
Due to the conservation of the total angular momentum component along the $z$ axis, the average nuclear spin decreases by the same value $\braket{\delta S_z}$. Thus, the nuclear spin relaxation time (by electrons) $T_{1,N}^e$ is determined by the relation
\begin{equation}
  \frac{\braket{J_z}}{T_{1,N}^e}=\frac{\braket{\delta S_z}}{\tau_R}.
\end{equation}
Substituting here Eq.~\eqref{eq:delta_S_ans} and comparing this with Eq.~\eqref{eq:T1e} we obtain
\begin{equation}
  \label{eq:relation}
  T_{1,N}^e=\frac{N_I}{f}T_{1,e}^N.
\end{equation}
Finally, the nuclear Zeeman reservoir heating rate is given by
\begin{equation}
  q_Z=\frac{N\braket{I_z}}{T_{1,N}^e}\hbar\gamma_N B.
\end{equation}
With the help of Eq.~\eqref{eq:relation}, it gives Eq.~\eqref{eq:q_Z} in the main text, which was obtained there from the general considerations.

\subsection{Spin-spin reservoir}

Now let us calculate the heating rate $q_{SS}$ of the nuclear spin-spin reservoir. Since Eq.~\eqref{eq:En} vanishes in the first order in $\mathcal H_{SS}$, it is necessary to consider the second order. It reads
\begin{multline}
  \label{eq:ESS_general}
  E_{SS}(t)=-\beta\Tr\Bigg\{\mathcal H_{SS}\mathcal H_{SS}(t)\\
  +\frac{\i}{\hbar}\int\limits_0^t\d t'\left(\mathcal H_{SS}\left[\mathcal H_{SS}(t'),\mathcal H_Z(t)\right]+\left[\mathcal H_Z,\mathcal H_{SS}(t')\right]\mathcal H_{SS}(t)\right)\\
  +\frac{1}{\hbar^2}\int\limits_0^t\d t'\int\limits_{t'}^t\d t''\left[\mathcal H_Z,\mathcal H_{SS}(t')\right]\left[\mathcal H_Z(t),\mathcal H_{SS}(t'')\right]\Bigg\},
\end{multline}
where the time dependence represents the evolution under action of the Hamiltonian $\mathcal H_0=\mathcal H_Z+\hbar\gamma_eBS_z+\mathcal H_{hf}$.

At the short time scales $t\ll\hbar/a$, the nuclear spin rotation angles are small so one can consider the fourth order in $\mathcal H$ only. One can check readily that this is equivalent to
\begin{multline}
  \label{eq:ESS_short}
  E_{SS}(t)=-\beta\Tr\left\{\mathcal H_{SS}\mathcal T\exp\left[\frac{\i}{\hbar}\int\limits_0^t\mathcal H_{hf}(t')\d t'\right]
  \right.\\\left.
    \times\mathcal H_{SS}\overline{\mathcal T}\exp\left[-\frac{\i}{\hbar}\int\limits_0^t\mathcal H_{hf}(t')\d t'\right]\right\},
\end{multline}
where $\mathcal T$ and $\overline{\mathcal T}$ denote time ordering and its reverse. This equation shows that for the calculation of the nuclear spin-spin energy, one can consider only the nuclear spin dynamics under $\mathcal H_{hf}(t)$, without additional spin precession due to the Zeeman term. Physically this is related to the facts that $\mathcal H_Z$ enters already the steady state density matrix~\eqref{eq:rho0_S} and that the small nuclear spin rotation angles commute.

Substituting Eq.~\eqref{eq:S0} in Eq.~\eqref{eq:ESS_short} we obtain
\begin{equation}
  \label{eq:delta_ESS}
  \braket{\Delta E_{SS}}=\frac{\beta(\hbar\gamma_N\tilde B_L\tau_c)^2}{24T_{2,e}^*}\left(1+\braket{\frac{2}{1+\Omega_e^2\tau_c^2}}\right),
\end{equation}
where
\begin{equation}
  {\tilde B}_L^2=-\frac{2}{N_I(\hbar\gamma_N)^2}\Tr\left[\bm J,\mathcal H_{SS}\right]^2.
\end{equation}
For dipole-dipole interaction, it equals $\sqrt{3}B_L$~\cite{S_abragam,S_OptOrCh2}, where the local nuclear field $B_L$ is defined by
\begin{equation}
  B_L^2=\frac{4}{N_I(\hbar\gamma_N)^2}\Tr\mathcal H_{SS}^2.
\end{equation}
From comparison of Eq.~\eqref{eq:delta_ESS} and Eq.~\eqref{eq:q_SS} in the main text, we obtain Eq.~\eqref{eq:Phi_ans} in the main text.

Calculation of the change of the nuclear spin-spin energy for long times $t\gtrsim\hbar/a$ after Eq.~\eqref{eq:ESS_general} is difficult. But neglecting the nuclear Zeeman splitting in the spin dynamics, one arrives again at Eq.~\eqref{eq:ESS_short}, which generally describes rotation of all nuclear spins by the same angle $\varphi$ around some axis. Then, we obtain the corresponding nuclear spin-spin energy
\begin{equation}
  \label{eq:ESS_phi}
  E_{SS}=-\frac{N_I\beta}{20}\left[1+2\cos(\varphi)+2\cos(2\varphi)\right]\left(\hbar\gamma_NB_L\right)^2.
\end{equation}
For short correlation time, one has $\phi=at/(2\hbar)$ and Eq.~\eqref{eq:ESS_phi} reproduces Eq.~\eqref{eq:delta_ESS} with $\Omega_e\tau_c\gg 1$. By contrast, for long correlation time, when $at/(2\hbar)\gg 1$ one can neglect the oscillating terms in Eq.~\eqref{eq:ESS_phi}, which gives
\begin{equation}
  \label{eq:Phi_long}
  \Phi=\frac{4}{5}\left(\frac{\hbar}{a\tau_c}\right)^2
\end{equation}
in a qualitative agreement with the general considerations in the main text.

\begin{figure*}
  \includegraphics[width=0.49\linewidth]{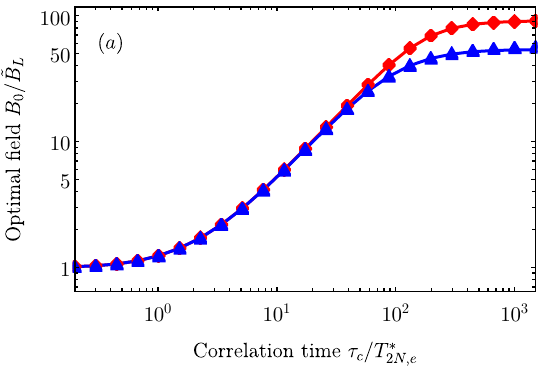}
  \includegraphics[width=0.49\linewidth]{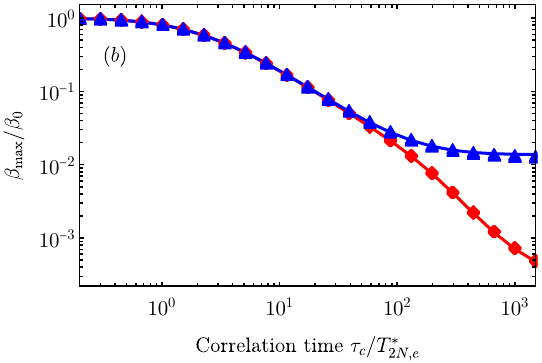}
  \caption{(a) Optimal magnetic field $B_0$ at which the maximum of the inverse nuclear spin temperature $\beta_{\rm max}$ is reached as a function of the correlation time $\tau_c$. (b) $\beta_{\rm max}$ as a function of $\tau_c$. The solid lines are calculated after Eq.~\eqref{eq:beta} in the main text using Eq.~\eqref{eq:phi_simple}, they are the same as in Fig.~\ref{fig:combo}(c,d) in the main text. The dots are calculated accounting for the electron-nuclear spin dynamics described by Eq.~\eqref{eq:omega_n}. Red and blue colors correspond to $N_I=10^6$ and $10^4$, respectively.
  }
  \label{fig:rot}
\end{figure*}

Another exact result can be found for the case of $\Omega_BT_{2,e}^*\gg1$, when external magnetic field is much stronger than the fluctuations of the Overhauser field (average Overhauser field can be added to $\Omega_B$ here). Then Eq.~\eqref{eq:ESS_general} yields
\begin{multline}
  \label{eq:ESS_complicated}
  E_{SS}(t)=-\beta\Tr\Big[\mathcal H_{SS}^2-\frac{1}{2}\sum_\pm\left(\frac{\omega_e}{\omega_e\pm\omega_B}\right)^2\Big(\mathcal H_{SS}^2\\
  -\mathcal H_{SS}\e^{\i(\omega_e\pm\omega_B)J_zt}\mathcal H_{SS}\e^{-\i(\omega_e\pm\omega_B)J_zt}\Big)\Big],
\end{multline}
where $\omega_e=a/(2\hbar)$ and $\omega_B=-\gamma_NB$ are the nuclear spin precession frequencies in the Knight and external magnetic fields, respectively. One can see that for $\omega_B=0$ this agrees again with Eqs.~\eqref{eq:ESS_short} and~\eqref{eq:ESS_phi}, where $\phi=at/(2\hbar)$. Then its average gives
\begin{equation}
  \Phi=\frac{2}{5}\left(\frac{\hbar}{a\tau_c}\right)^2\left(2-\frac{1}{1+(a\tau_c/\hbar)^2}-\frac{4}{4+(a\tau_c/\hbar)^2}\right).
\end{equation}
This expression represents an interpolation between Eq.~\eqref{eq:Phi_ans} in the main text and Eq.~\eqref{eq:Phi_long}.

Fig.~\ref{fig:rot} shows comparison of the simplified calculation with
\begin{multline}
  \label{eq:phi_simple}
  \Phi=
  \left<\frac{1}{1+\Omega_e^2\tau_c^2}\right>+\\
  \frac{2}{5}\left(\frac{\hbar}{a\tau_c}\right)^2\left(2-\frac{1}{1+(a\tau_c/\hbar)^2}-\frac{4}{4+(a\tau_c/\hbar)^2}\right),
\end{multline}
which was shown in Fig.~\ref{fig:combo} in the main text (solid lines) and exact calculation based on Eq.~\eqref{eq:ESS_phi} (dots). In the latter case, the nuclear spin dynamics was described using an exact solution obtained in Ref.~\onlinecite{S_PhysRevLett.126.216804}. It represents rotation of average electron and nuclear spins around external magnetic field with the frequency
\begin{equation}
  \label{eq:omega_n}
  \omega_n=\omega_e\frac{\Omega_B}{\Omega_e}
\end{equation}
neglecting the nuclear Zeeman splitting ($\omega_B=0$). One can see almost perfect agreement between the two calculations.

In the opposite limit of $\omega_B\gg\omega_e$ and large nuclear spin rotation angles, the change of the nuclear spin-spin energy in Eq.~\eqref{eq:ESS_complicated} is suppressed by a factor $(\omega_e/\omega_B)^2$ because of the contribution of $\mathcal H_Z$ to the initial nuclear spin density matrix~\eqref{eq:rho0_S}. Depending on the electron localization strength and nuclear gyromagnetic ratio, this may lead to appearance of a second maximum in the dependence $\beta(B)$ at higher magnetic fields, when heating of the nuclear spin-spin reservoir gets suppressed.

\section{S2. Comparison of $T_{2,N}$ and $T_{1,N}^e$}

Establishment of the nuclear spin temperature requires the thermalization time $T_{2,N}$ to be shorter than the nuclear spin polarization time by localized electrons $T_{1,N}^e$ given by Eq.~\eqref{eq:relation}. The latter increases with increase of magnetic field and nonmonotonously depends on the correlation time $\tau_c$. It reaches the minimum of approximately $4.9N_IT_{2,e}^*/f$ at $\tau_c\approx0.88T_{2,e}^*$. In GaAs-based quantum dots, typically $aN\sim100~\mu$eV and $T_{2,N}\sim0.1$~ms. Therefore the nuclear spin temperatures establishes at $N^{3/2}/f>10^7$. This condition is always satisfied for $N=10^6$, while for $N=10^4$ it requires $f<0.1$. This shows that the nuclear spin temperature is established in most of experiments.

The strong magnetic field or Overhauser field may prevent thermalization, but experiments with quantum dots do not show this~\cite{S_Chekhovich_constants,S_PhysRevB.97.041301}, probably due to the inhomogeneity of the hyperfine interaction and strain.

\end{document}